\documentclass[journal=jacsat,manuscript=article]{achemso}
\usepackage{graphicx}
\usepackage{subcaption}
\usepackage{booktabs}
\usepackage{siunitx}
\usepackage[dvipsnames]{xcolor}
\usepackage{amssymb}
\usepackage{pifont}
\usepackage{wrapfig}
\usepackage{hyperref}

\usepackage{amsmath} 

\usepackage{tabularx,colortbl}
\usepackage{capt-of}
\usepackage{adjustbox}
\usepackage[font=normalsize,labelfont={bf},font={bf}]{caption}

\author{Desmond Kuan}
\affiliation[biomed]
{Department of Biomedical Engineering, Carnegie Mellon University, 15213, USA}

\author{Amir Barati Farimani}
\email{barati@cmu.edu}
\affiliation[meche]
{Department of Mechanical Engineering, Carnegie Mellon University, 15213, USA}
\alsoaffiliation[biomed]
{Department of Biomedical Engineering, Carnegie Mellon University, 15213, USA}
\alsoaffiliation[mld]
{Machine Learning Department, Carnegie Mellon University, 15213, USA}

\title[An \textsf{achemso} demo]
{AbGPT: De Novo Antibody Design via Generative Language Modeling}

\begin{document}

\begin{abstract}

The adaptive immune response, largely mediated by B-cell receptors (BCRs), plays a crucial role for effective pathogen neutralization due to its diversity and antigen specificity. Designing BCRs de novo, or from scratch, has been challenging because of their complex structure and diverse binding requirements. Protein language models (PLMs) have shown remarkable performance in contextualizing and performing various downstream tasks without relying on structural information. However, these models often lack a comprehensive understanding of the entire protein space, which limits their application in antibody design. In this study, we introduce Antibody Generative Pretrained Transformer (AbGPT), a model fine-tuned from a foundational PLM to enable a more informed design of BCR sequences. Using a custom generation and filtering pipeline, AbGPT successfully generated a high-quality library of 15,000 BCR sequences, demonstrating a strong understanding of the intrinsic variability and conserved regions within the antibody repertoire.

\noindent  \href{https://github.com/deskk/AbGPT}{https://github.com/deskk/AbGPT}

\end{abstract}

\section{Introduction}
B-cell receptors (BCRs) are essential components of the adaptive immune system, recognizing and neutralizing antigens through their highly variable antigen-binding regions, generated via somatic recombination. This process creates a diverse antibody repertoire capable of recognizing millions of antigens, making BCRs and their monoclonal antibody (mAb) derivatives a focal point in therapeutic research for cancer, autoimmune disorders, and infectious diseases due to their high specificity and low toxicity. \cite{alberts2002b,FISCHER2018176,janeway2001immunobiology,abraham2016structure,christian2016ecm} However, traditional methods of developing therapeutic antibodies, such as phage display, are often slow, expensive, and limited in scope. \cite{zambrano2022high} Somatic hypermutation adds to the complexity, introducing further diversity in the BCR repertoire by mutating the variable regions at a high rate. The limitations of traditional approaches have prompted a shift toward computational methods, such as sequence-based design, which avoids the need for explicit structural data and instead uses the primary amino acid sequence to predict functional properties. Structure-based methods, while effective, rely on detailed 3D structures of antibodies and antigens, which are often difficult to obtain and model accurately due to the resource-intensive nature of experimental methods like X-ray crystallography and cryo-electron microscopy. \cite{farimani2016computational, wang2023graph,barati2017antibody,murata2018cryo,fernandez2023challenges} In contrast, sequence-based methods offer a more flexible alternative by focusing directly on genetic information.

Large language models (LLMs), which have revolutionized natural language processing (NLP), provide a robust foundation for advancing sequence-based design. \cite{radford2019better} Originally developed to understand and generate human language, LLMs have demonstrated exceptional capabilities in learning from sequential data. These models, including prominent examples such as the Transformer architecture, Bidirectional Encoder Representations from Transformer (BERT), and Generative Pretrained Transformer (GPT), have set new benchmarks in NLP by capturing complex patterns, dependencies, and contextual relationships within textual data. \cite{vaswani2017attention, devlin2018bert, radford2018improving} The success of LLMs in NLP has inspired their application to other domains where sequential data plays a critical role, including protein science. \cite{valentini2023promises} Protein language models (PLMs) build on these concepts by training on large datasets of protein sequences, identifying patterns and motifs relevant to structure and function. \cite{ofer2021language,rao2019evaluating,evans2021protein} PLMs can generate rich sequence representations applicable to tasks like 3D structure prediction, binding affinity estimation, and protein-protein interaction analysis. \cite{madani2023large,van2022foldseek,bhat2023novo,cong2019protein,su2023saprot,alley2019unified,kim2024gpcr,badrinarayanan2024multi,guntuboina2023peptidebert,mollaei2024idp} 

The success of PLMs has led to the development of several foundation models that have significantly advanced our understanding of protein science. Models such as ESM and ProtTrans have demonstrated the ability to predict and contextualize the complex dynamics of proteins, offering insights that were previously unattainable. \cite{rives2021biological,elnaggar2021prottrans} In the field of antibody research, models like AntiBERTa \cite{leem2022deciphering} have leveraged the RoBERTa architecture to pre-train and contextualize BCR sequence repertoire, achieving notable performance in tasks such as paratope predictions. Similarly, ProGen2-OAS \cite{nijkamp2023progen2} has been trained on a vast dataset of unpaired antibody sequences to generate viable antibody sequences, while IgLM \cite{shuai2023iglm} has been developed to facilitate controlled generation and infilling of specific antibody sequences through the use of conditional tags for species and chain types. Additionally, other models have incorporated native pairing of heavy and light antibody chains in their training to capture inter-chain features effectively. \cite{burbach2024improving} However, most antibody-specific models are limited by their focus on antibody-specific data, often overlooking the broader principles governing the entire protein space. While this helps address the challenge of sequence hypervariability in antibody complementarity-determining regions (CDRs), it also restricts these models from accessing the comprehensive insights available from foundational PLMs, which have been trained on the whole protein space and understand the fundamental principles of protein science. Recent work by Singh et al. \cite{singh2023learning} demonstrated the effectiveness of a transfer learning approach that leverages foundational PLMs, yielding remarkable results for CDR design by incorporating a supervised learning approach that is trained on antibody structure and binding specificity profiles. This highlights the potential of foundational PLMs in striking a balance between broad generalization and task-specific optimization.

Inspired by this concept, our approach, Antibody Generative Pretrained Transformer (AbGPT), is specifically designed to overcome the limitations associated with existing methods in the de novo design of antibody sequences. Building on foundational PLMs, AbGPT directly tackles the challenges associated with sequence hypervariability and the constraints of traditional structure-based design. The model was fine-tuned on a comprehensive dataset of 71.98 million unique BCR sequences, spanning 61 studies, previously used in AntiBERTa's work. The dataset includes 52.89M unpaired heavy chains and 19.09M unpaired light chains. By fine-tuning ProtGPT2 \cite{ferruz2022protgpt2} with these natural BCR sequences, AbGPT captures the richness of representations and unique hypervariability inherent in the repertoire. AbGPT leverages the generalized insights from foundational PLMs while maintaining a deep focus on the BCR-specific features. This allows AbGPT to explore a wider design space, generating novel BCR sequences with desirable, yet rare, mutations that traditional methods may overlook. As a result, AbGPT expands the synthetic BCR library and opens up new possibilities for antibody discovery and optimization. By integrating large-scale sequence data with generative language modeling, AbGPT represents a more scalable and versatile framework for therapeutic antibody design.

\section{Methods}
\subsection{Overview of AbGPT}
\begin{figure}[t!]
     \centering
     \includegraphics[width=1.0\linewidth]{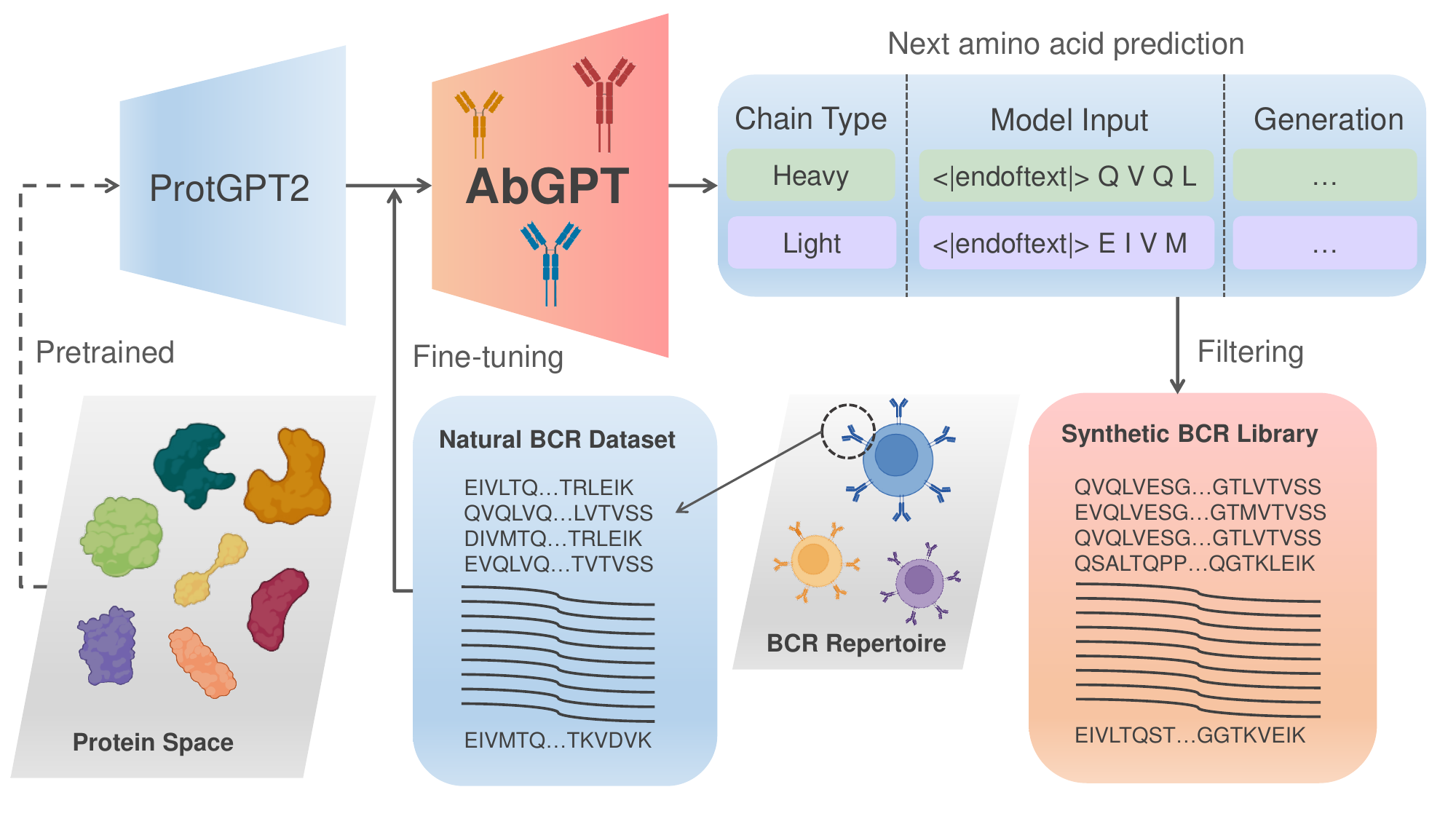}
     \caption{The overview of AbGPT training and sequence generation. Given that ProtGPT2 was trained with the entire protein space, we fine-tuned this foundational model with natural BCR sequences to capture the rich representations of BCR while retaining the fundamental principles in protein space. We utilize a novel generation and filtering framework to create a high-quality synthetic BCR library.} 
     \label{fig:model}
 \end{figure}
 
Similar to ProtGPT2, AbGPT utilizes a GPT-2 Transformer decoder architecture, which is an autoregressive generative model, to specifically design novel BCR sequences. An autoregressive model is trained with causal language modeling, where predictions are made in a sequential manner, using only the previous observations to predict the next token. \cite{shin2021protein} In the context of antibody sequences, the model predicts each amino acid by conditioning on the previously generated portion of the sequence. Each prediction step uses the generated part of the sequence that precedes it to inform the next prediction. Specifically, the probability of a particular amino acid ($a_i$) depends solely on its context, which consists of the preceding tokens in the sequence ($a_{<i}$). The total probability of generating a complete sequence ($A$) is thus the product of the individual probabilities for each amino acid:

\begin{equation}
p(A) = \prod_{i=1}^{n} p(a_i \mid a_{<i})
\end{equation}

This autoregressive approach is particularly advantageous for BCR sequence generation because it allows the model to capture complex dependencies between amino acids, including the CDRs and FRs of a BCR sequence. These dependencies are critical for maintaining the structural integrity and functional specificity of antibodies, especially within the highly variable CDR regions. By considering the sequence context at each step, the model can generate BCRs that are more likely to retain essential biochemical properties and functional capabilities. AbGPT is designed to capture both the fundamental principles of protein science and the unique hypervariability of antibody CDRs. We aim to utilize this dual focus to ensure the model captures general protein structures and also designs BCRs with diversified and viable antigen-binding properties. The model is trained similarly to ProtGPT2 by minimizing the negative log-likelihood over the entire dataset. \cite{ferruz2022protgpt2} This training objective ensures that the model learns to accurately predict the conditional probability of the next amino acid given the preceding sequences, and this is done for each sequence $k$ in the dataset ($D$). Formally, the loss function for this training is expressed as:

\begin{equation}
\mathcal{L}_{\text{CLM}} = -\sum_{k=1}^{D} \log p_{\theta} \left( a_{i}^{k} \mid a_{<i}^{k} \right)
\end{equation}

This step-by-step approach ensures that the dependencies between different parts of the sequence are captured which is crucial for maintaining the structural and functional properties of the antibodies.

\subsection{Data processing}
In our efforts to enhance the diversity of human BCR sequence generation and improve model performance, we adapted AntiBERTa’s pretraining dataset for fine-tuning. \cite{leem2022deciphering} This dataset is drawn from a collection of 61 individual studies curated from Observatory Antibody Space (OAS). \cite{kovaltsuk2018observed,olsen2022observed} The human BCR sequences in the dataset were processed to remove errors and applied specific length constraints, collecting a total of 71.98 million unique human BCR sequences. The dataset was divided into training, validation, and testing subsets, maintaining proportions of 8:1:1. For our purposes, we merged the training and validation datasets into a single set, while retaining the original testing dataset, resulting in a final ratio of 9:1. To prepare the dataset for fine-tuning ProtGPT2, each sequence was formatted with 60 amino acids per line in the FASTA format, and the FASTA format header with “$<|endoftext|>$” token to standardize the input format for the autoregressive model. \cite{radford2019language}

\subsection{Generation and filtering framework}

The objective of our generation pipeline is to create a diverse and functional synthetic BCR library by leveraging chain-specific starting prompts for our generative model. We provide 4 starting amino acid residues as context prompting to discern between heavy and light chain types during generation. We identified the most prevalent starting amino acid combinations for both light and heavy chains. For the initial synthetic BCR library generation, we then picked the 5 most frequent starting residues of 4 amino acids each from heavy and light chain types observed in the training dataset to generate BCR sequences. This approach is designed to help the model recognize and differentiate between chain types, as BCR heavy and light chains typically exhibit different starting residues. \cite{janeway2001immunobiology} We experimented with hyperparameters such as top-k sampling and repetition penalty. Top-k controls the next token selection to the top-k most probable tokens at each step and offers a middle ground between creativity and exploration. A large k-value increases the diversity of word selection by considering a broader array of tokens. The repetition penalty discourages the model from generating repeated tokens. Analyzing the length distribution of our training dataset, we observed a distinct bimodal distribution pattern, which we hypothesize corresponds to the distributions of heavy and light chains in the BCR repertoire. Given this observation, and recognizing that each token in AbGPT represents approximately four amino acids, we imposed length restrictions on the generation process for each chain type. Specifically, we limit the minimum generation length to 28 tokens (approximately 128 amino acids) for heavy chains and 20 tokens (approximately 80 amino acids) for light chains. Post-generation, the sequences were screened to meet specific length constraints: light chains were retained if they were between 100 and 120 amino acids, and heavy chains if they were between 110 and 140 amino acids. To further ensure the quality of our synthetic BCR sequence library, we applied a perplexity-based filtering step. Perplexity (\text{ppl}) is defined as the exponentiated average negative log-likelihood, is given by \( \mathbf{x} = (x_0, x_1, \dots, x_t) \), then the perplexity of \( \mathbf{x} \) is:

\begin{equation}
\text{ppl}(\mathbf{x}) = \exp\left\{ -\frac{1}{t} \sum_{i=1}^{t} \log p_{\theta}(x_i \mid x_{<i}) \right\}
\end{equation}

where \( \log p_{\theta}(x_i \mid x_{<i}) \) is the log-likelihood of the \( i \)-th token conditioned on the preceding tokens \( x_{<i} \) according to our model. This metric evaluates the model’s ability to predict tokens in a sequence. Although there is no universally established threshold for delineating the quality of the sequences, we set the threshold perplexity value at 13.0 or below based on our experimentation. We expect this comprehensive pipeline to increase the likelihood of generating a functional and diverse synthetic BCR library, which in turn will generate novel sequences that are more likely to exhibit similar properties to those of natural BCRs.

\subsection{Solubility and aggregation propensity}

mAbs, the secreted form of BCRs, are known to aggregate under non-native conditions, which can lead to a loss of functionality and increased toxicity. Therefore, evaluating the solubility and aggregation propensity of the generated BCR sequences is crucial for the developability of mAbs. Since the antigen-binding site of BCR shares the same antigen specificity as the mAbs that the B cell will eventually produce \cite{janeway2001antigen}, we focused our in-silico analysis on the most variable region for its functions, which is the Complementarity-Determining Region 3 of the heavy chain (CDR-H3).  To accurately identify the CDR-H3 region in each BCR sequence, we employed the Antibody Numbering and Antigen Receptor ClassIfication (ANARCI) tool \cite{dunbar2016anarci}, which assigns residues according to the standardized ImMunoGeneTics (IMGT) numbering scheme. The IMGT system is a standardized numbering scheme for annotating variable domains of immunoglobulins. \cite{lefranc2009imgt} To measure the solubility of the generated sequences, we used the CamSol method for protein solubility prediction. \cite{sormanni2015camsol} Camsol Intrinsic is a sequence-based predictor that provides rapid assessments of intrinsic solubility profiles and overall solubility scores. CamSol assigns scores to each residue based on its physicochemical properties, measuring the impact of individual residues on the protein's overall solubility. In addition to solubility, we evaluated the average aggregation propensity values per amino acid of the generated BCR sequences using AGGRESCAN. \cite{conchillo2007aggrescan} AGGRESCAN measures the overall natural tendency of the sequences to aggregate.

\subsection{Structural stability}
The ability to design BCR sequences that fold into stable, ordered structures is a key indicator of the generative model's understanding of protein folding principles. Evaluating the structural stability of the de novo BCR sequences generated by AbGPT is crucial for assessing their potential functional efficacy. Antibody sequence design is particularly challenging due to the presence of hypervariable CDRs, which help in antigen binding. Accurate prediction and maintenance of structural integrity in these regions are essential for the design of functional antibodies. To address these challenges, we employed AlphaFold2 \cite{jumper2021highly,evans2021protein} for structure predictions. For each generated sequence, we collected the predicted local distance difference test (pLDDT) score, with a higher score indicating greater confidence in the structural predictions. Importantly, the pLDDT score has been shown to correlate strongly with the structural order of proteins, making it a valuable metric for assessing the stability and proper folding of the generated sequences. \cite{ferruz2022protgpt2} We obtained pLDDT scores for both full-length sequence generation and their standalone CDR-H3 regions, annotated using the IMGT numbering scheme. Additionally, we visualize the 3D structures of selected sequences that share similar framework regions using PyMOL. This approach allowed us to observe how the model handles the hypervariability in CDR-H3, providing insights into its ability to generate structurally stable sequences.

\subsection{Diversity}
The diversity of natural BCR repertoire arises mainly from somatic hypermutation and antibody germline mutation of variable (V), diversity (D), and joining (J) gene segments. \cite{victora2022germinal,eisen1964variations,eisen2014affinity,mcheyzer2012molecular} Germline mutation plays an important role in the diversity of antibodies, these mutations occur in the variable regions of antibody genes before antigen exposure. We wanted to study if the model is capable of generating BCR sequences that are novel and diverse yet not too dissimilar from the real BCR sequences. We first grouped them by the V and J gene combinations using ANARCI germline assignment and performed CDR-H3 sequence length distribution analysis. Then, we performed t-distributed stochastic neighbor embedding (t-SNE) to visualize the embedding distribution of the generated sequences compared to the training data. We aim to answer the question of whether the model is only generating sequences highly similar to training data or if it knows how to 'design' novel sequences while staying relatively reasonable within the boundary of similar properties to human antibody sequences. To investigate further, we also generated the t-SNE plot grouped by different V-gene families, as well as the V-gene gene segment within a family.

\subsection{Humanness evaluation}

One of the therapeutically relevant properties of antibodies is the assessment of the human-likeness of generated antibodies, as it is crucial for evaluating the ‘naturalness’ of the antibody design and minimizing the risk of provoking an immune response. \cite{schmitz2020human} To assess the humanness of the generated BCRs, we utilized the BioPhi computational tool for naturalness assessment. \cite{prihoda2022biophi} BioPhi is an evaluation tool designed to calculate the prevalence of antibody sequences across the human population. By evaluating the generated BCR sequences via natural antibody repertoires and germline sequence identity, we were able to obtain an interpretable humanness score for each sequence. This allows us to examine whether AbGPT can produce BCR sequences that would closely resemble naturally occurring human antibodies. 

\section{Results}
\subsection{AbGPT for BCR sequence design}

Recent advancements in antibody design using generative language models have marked significant progress in the field. However, to the best of our knowledge, current approaches are limited by focusing exclusively on models trained within the antibody space. This narrow focus overlooks the potential benefits of leveraging the broader protein space, which contains fundamental principles that could enhance antibody design. We argue that this limitation may restrict the diversity and efficacy of the generated BCR sequences. To address the limitation, we introduce AbGPT, a generative language model capable of designing full-length BCR sequences that have high developability, stability, and desirable therapeutic properties. Our approach utilizes a large PLM (ProtGPT2) that exhibits a strong foundational understanding of the whole protein space and fine-tunes it with a natural BCR dataset. ProtGPT2 was pretrained with approximately 50M sequences spanning the entire protein space and capable of generating sequences that are similar yet distant to natural BCR sequences. We fine-tuned ProtGPT2 with 57M human BCR sequence data containing heavy and light chain types. The fine-tuning step aimed to enhance the model's precision in designing an artificial human BCR library. Our goal is to expand the design capability of our selected approach by following the principles of natural proteins, while not being restricted to only the antibody space. With our generation and filtering pipeline, we successfully generated a synthetic BCR library of over 15,000 light and heavy chain sequences. This BCR repertoire will increase the likelihood of discovering unique antibodies with desirable therapeutic properties. 

\subsection{AbGPT designs soluble and non-aggregating BCR sequences}

In mAbs, aggregation can compromise their efficacy and safety as therapeutic agents. Similarly, for BCRs, aggregation can disrupt B cell function and signaling, which may subsequently affect the secretion of mAbs. The CDR regions of BCR are highly variable, and this hypervariability may introduce regions of hydrophobicity, which could increase the likelihood of aggregation if the sequence design does not adequately address this aspect. Therefore, it is crucial to ensure the sequence design process carefully considers the ideal biochemical properties of natural membrane immunoglobulins. 
\begin{figure}[h] 
    \centering
    \begin{subfigure}[t]{0.32\linewidth}
        \caption[]{}
        \includegraphics[width=\linewidth]{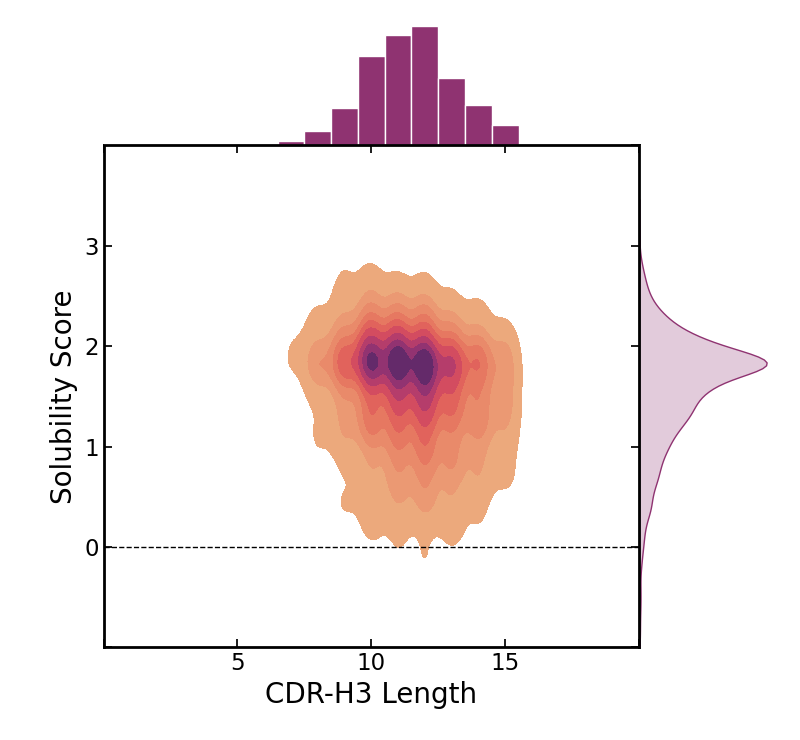}
        \label{fig:sol}
    \end{subfigure}
    \hfill
    \begin{subfigure}[t]{0.32\linewidth}
        \caption[]{}
        \includegraphics[width=\linewidth]{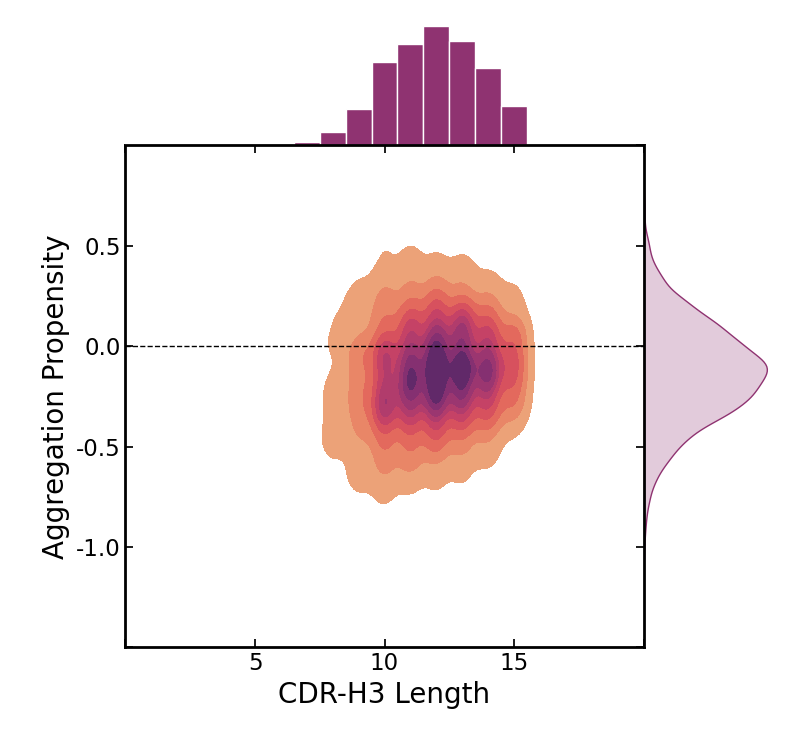}
        \label{fig:agg}
    \end{subfigure}
    \hfill
    \begin{subfigure}[t]{0.32\linewidth}
        \caption[]{}
        \includegraphics[width=\linewidth]{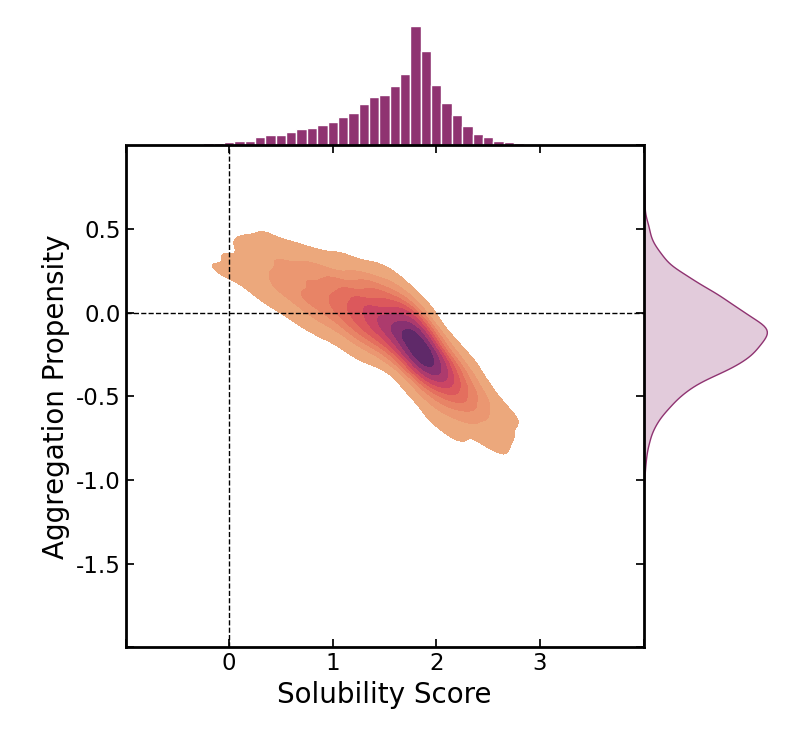}
        \label{fig:agg sol}
    \end{subfigure}
    \begin{subfigure}[t]{0.32\linewidth}
        \caption[]{}
        \includegraphics[width=\linewidth]{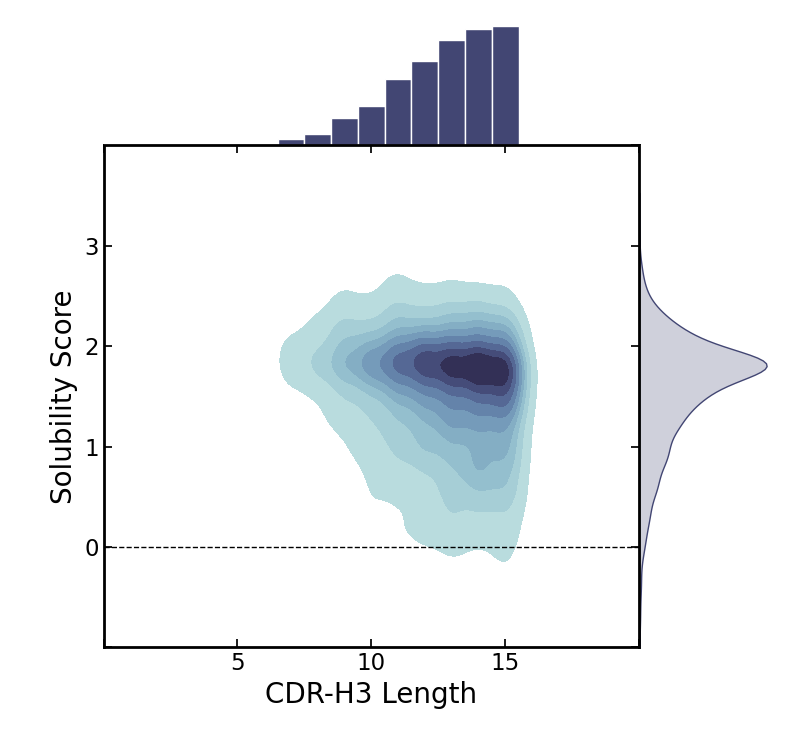}
        \label{fig:sol train}
    \end{subfigure}
    \hfill
    \begin{subfigure}[t]{0.32\linewidth}
        \caption[]{}
        \includegraphics[width=\linewidth]{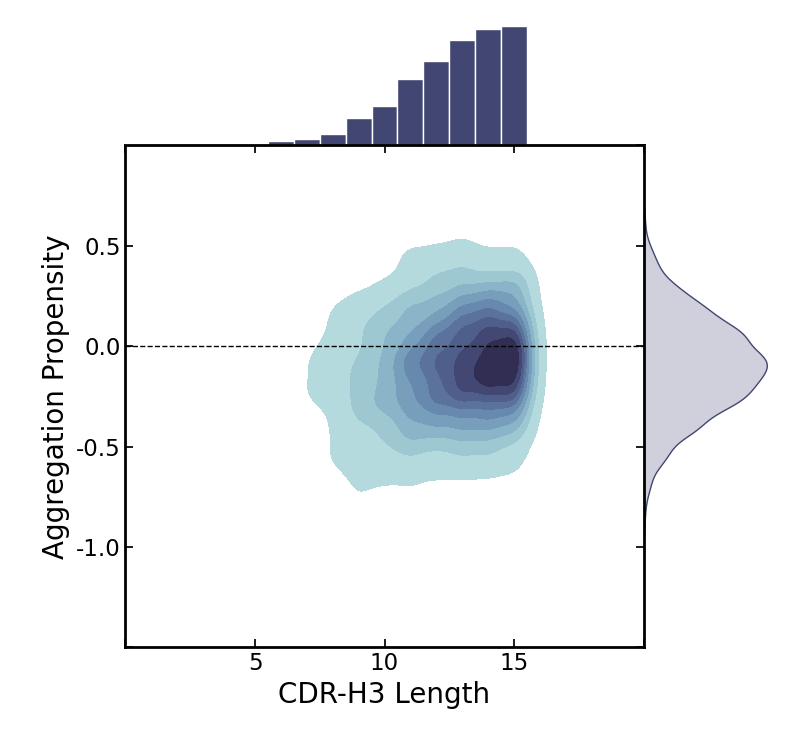}
        \label{fig:agg train}
    \end{subfigure}
    \hfill
    \begin{subfigure}[t]{0.32\linewidth}
        \caption[]{}
        \includegraphics[width=\linewidth]{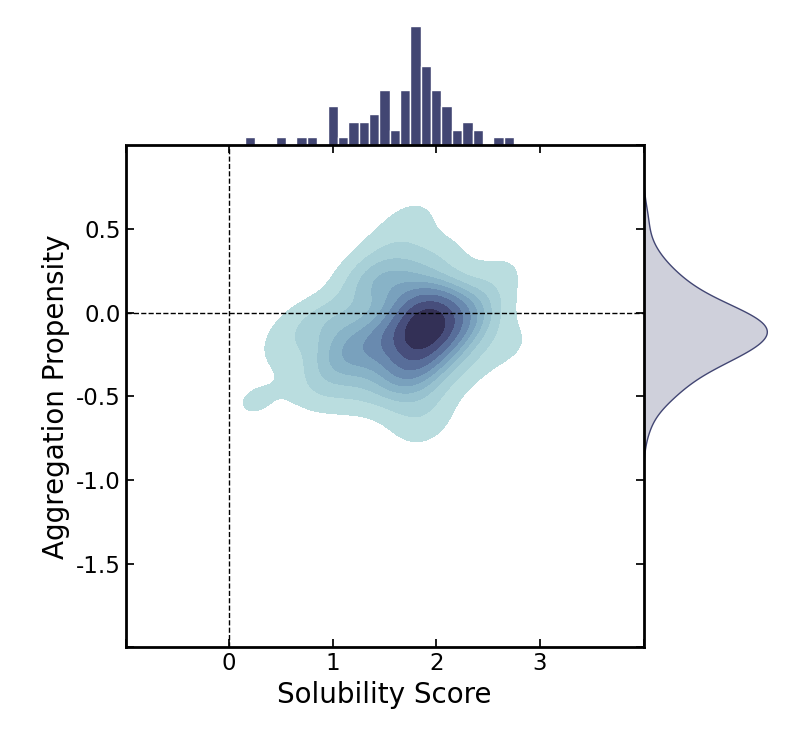}
        \label{fig:agg sol train}
    \end{subfigure}
    \caption{Biological properties of the CDR-H3 region comparing the generated BCR library and the natural BCR repertoire. (a, d) Solubility profiles by CDR-H3 length variations for the generated BCR library (a) and the natural BCR repertoire (d). (b, e) Aggregation propensity profiles by CDR-H3 length variations for the generated BCR library (b) and the natural BCR repertoire (e). (c, f) Relationship between solubility and aggregation profiles for the generated BCR library (c) and the natural BCR repertoire (f).}
    \label{fig:comparison}
\end{figure}
By analyzing the kernel density estimate (KDE) plot of solubility and aggregation propensity from Figure \ref{fig:sol}, we show that AbGPT is capable of producing a BCR library with high solubility profiles, which is crucial for their functional expression and stability. We observe a presence of clusters with high solubility profiles, with the most sequences concentrated between the lengths of 10 to 12, distinguishable by their darkest hue. High solubility scores indicate regions with residues that will significantly enhance the solubility of the antibodies. In addition, there is no direct correlation between solubility score and sequence length. In other words, the solubility is not affected by the variation in generation lengths. A similar pattern can be observed from Figure \ref{fig:agg} of average aggregation propensity and sequence length. This suggests that sequence length design can be explored without necessarily increasing the risk of aggregation, thus providing more flexibility in designing functional BCRs. From Figure \ref{fig:agg sol}, we observe a downward trend in the density plot, and this is expected, as solubility increases, aggregation propensity tends to decrease. These solubility and aggregation propensity profiles of the synthetic BCR library are comparable to the natural BCR repertoire in the training data. This demonstrates the model's excellent capability in designing synthetic BCRs with high developability qualities. Further screening on BCRs within the higher solubility and lower aggregation propensity 'hotspots' for experimental validation could streamline the development of highly functional BCRs. By comparing the biological properties of our generated BCR library (Figure \ref{fig:sol}-\ref{fig:agg sol train}) with the natural BCR repertoire from training data (Figure \ref{fig:sol train}-\ref{fig:agg sol train}), we notice differences in where great densities of high solubility and low aggregation propensity are located in both sets of plots. This is likely due to the length restrictions we impose during the sequence generation. However, this does not affect the model's capability to generate desirable properties comparable to the natural BCR repertoire.

\subsection{BCR sequences are structurally stable}

Another biophysical property that heavily influences the developability of the BCR sequence is structural stability. Structural stability refers to the ability of the antibody to maintain its three-dimensional structure under certain physiological conditions. Analyzing the structural information helps us understand the folding and the binding properties of these generated BCR sequences. We predicted the 3D structures of the generated sequences using Alphafold2 and reported the pLDDT score. pLDDT is the per-atom confidence estimate on a 0-100 scale, where a higher value corresponds to higher confidence in prediction. We compared the average pLDDT metric of both the full-length heavy chain and variable region CDR-H3. 
\begin{figure}[h] 
    \centering
    \begin{subfigure}[t]{0.495\linewidth}
        \caption[]{}
        \includegraphics[width=0.9\linewidth]{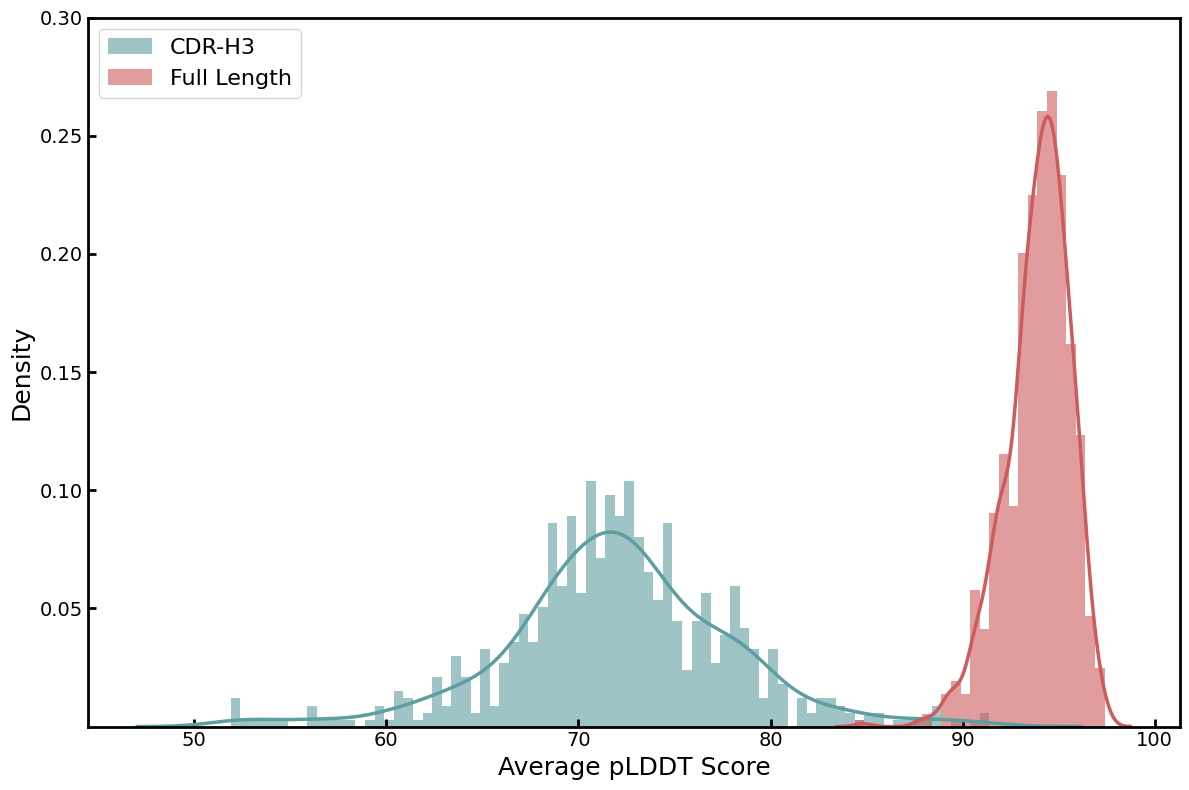}
        \label{fig:plddt}
    \end{subfigure}
    \hfill
    \begin{subfigure}[t]{0.495\linewidth}
        \caption[]{}
        \includegraphics[width=\linewidth]{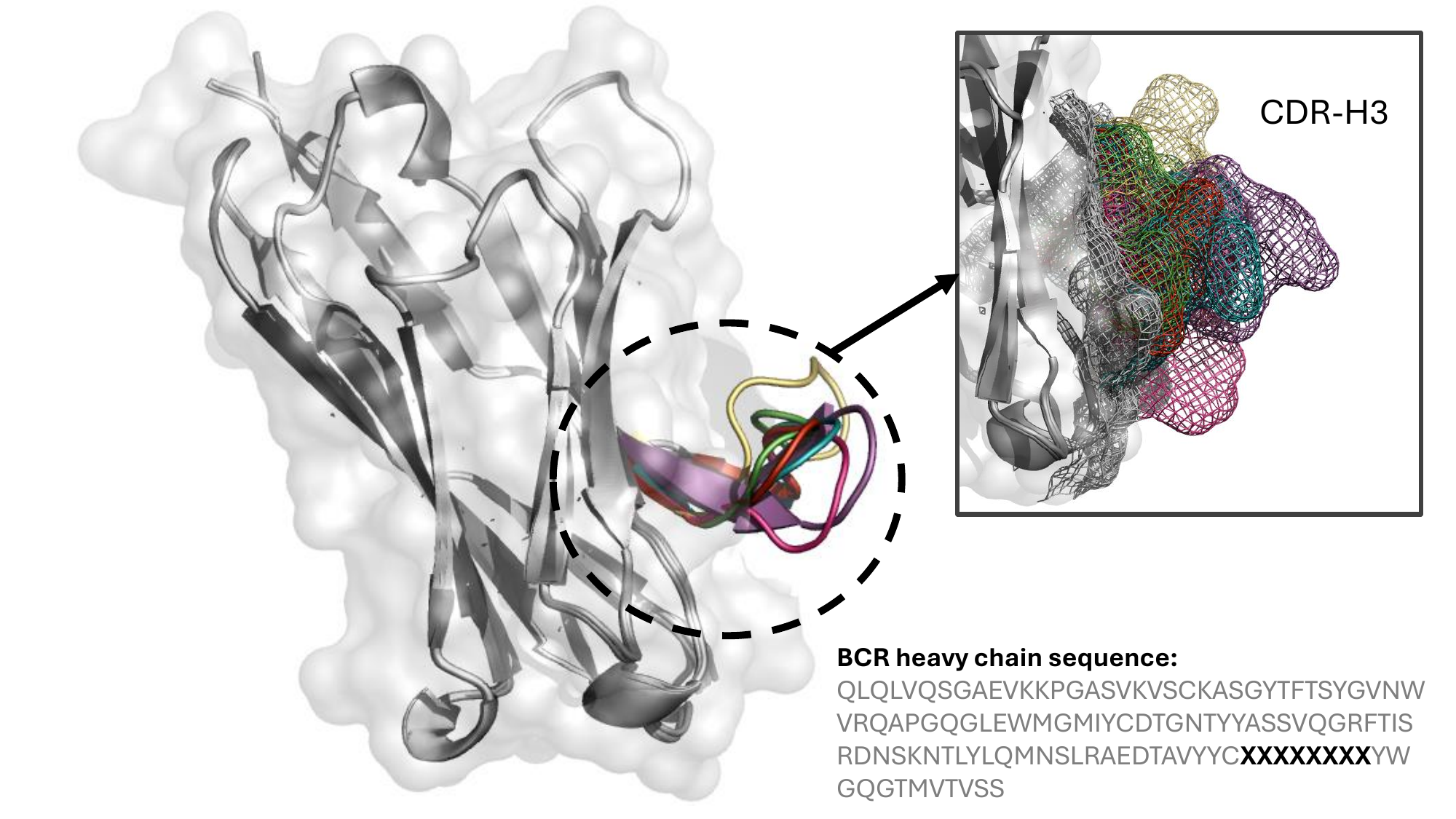}
        \label{fig:structure}
    \end{subfigure}
    \caption{The structural analysis of generated BCR sequences. (a) Average pLDDT score comparison of CDR-H3 and full-length heavy chain sequences. The average pLDDT score for CDR-H3 has a lower pLDDT score and a wider distribution compared to the full-length sequences. (b) Generated BCR heavy chain sequence structures visualized using PyMol that show different CDR-H3 configurations. A zoomed-in view of the CDR-H3 region with mesh surface visualization is included to show the variability in AbGPT generation. CDR-H3 is indicated as 'XXXXXXXX' and varied in length.}
    \label{fig:structural_analysis}
\end{figure}
From Figure \ref{fig:plddt}, the distribution of CDR-H3 shows a broader distribution compared to the full-length sequence, the average pLDDT score for the full-length sequences is significantly smaller and better than the CDR-H3 region. This shows that the model predictions for the full-length sequences are more confident, which is expected since the overall BCR structure is relatively more structured and less variable compared to the most variable region, CDR-H3 of the antibody sequence. This shows the model inherently knows which part of the BCRs needs to be kept fairly constant even without supervision during the full-length generation. On the other hand, the moderate pLDDT score of CDR-H3 reflects the intrinsic challenges in predicting the CDR-H3 regions, as the corresponding high-quality multiple sequence alignments (MSAs) are unavailable for antibodies in the AlphaFold model. \cite{singh2023learning} In Figure \ref{fig:structure}, we selected a few BCR structures predicted for 3D visualization on how the CDR-H3 regions can be varied using AbGPT. For this demonstration purpose, we selected sequences with high predictive scores for CDR-H3 ($\geq$ 85 \% pLDDT). Our findings highlight the effectiveness of AbGPT in discriminating between the highly mutable CDR regions and the FR that maintain the structural integrity of BCRs.


\subsection{AbGPT creates a diversified BCR repertoire and understands biological representations}

\begin{figure}[h] 
    \centering
    \begin{subfigure}[t]{0.495\linewidth}
        \caption[]{}
        \includegraphics[width=\linewidth]{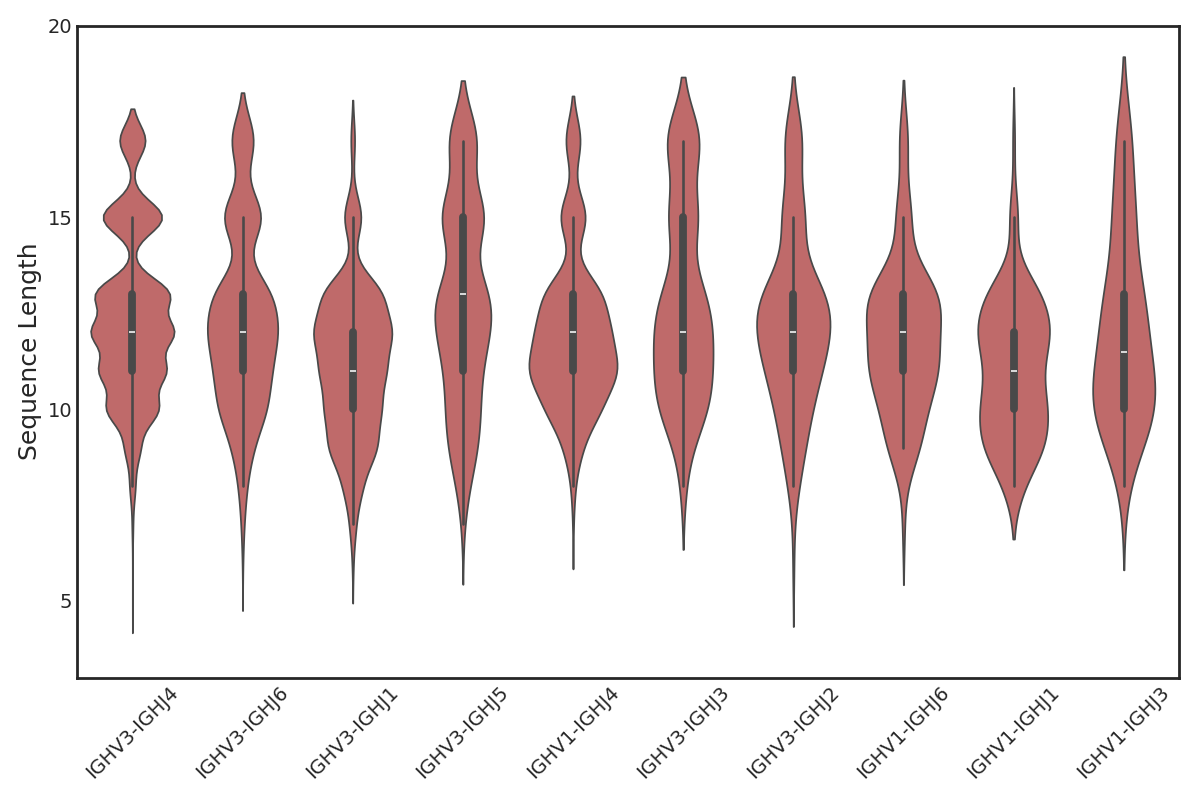}
        \label{fig:length dist}
    \end{subfigure}
    \hfill
    \begin{subfigure}[t]{0.495\linewidth}
        \caption[]{}
        \includegraphics[width=\linewidth]{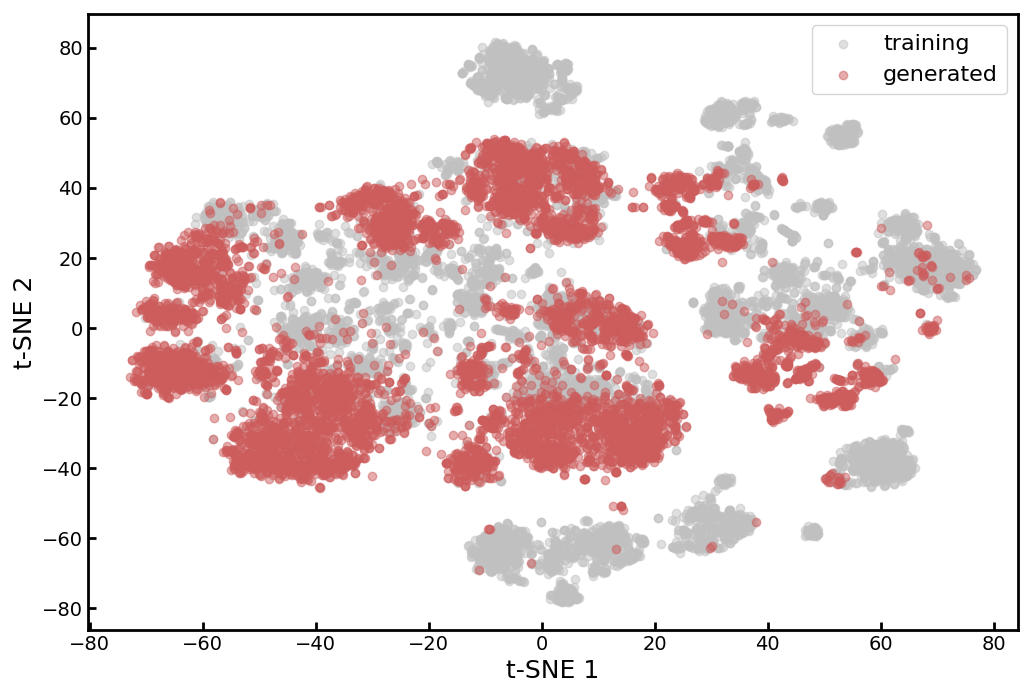}
        \label{fig:tsne train and gen}
    \end{subfigure}
    \hfill
    \begin{subfigure}[t]{0.495\linewidth}
        \caption[]{}
        \includegraphics[width=\linewidth]{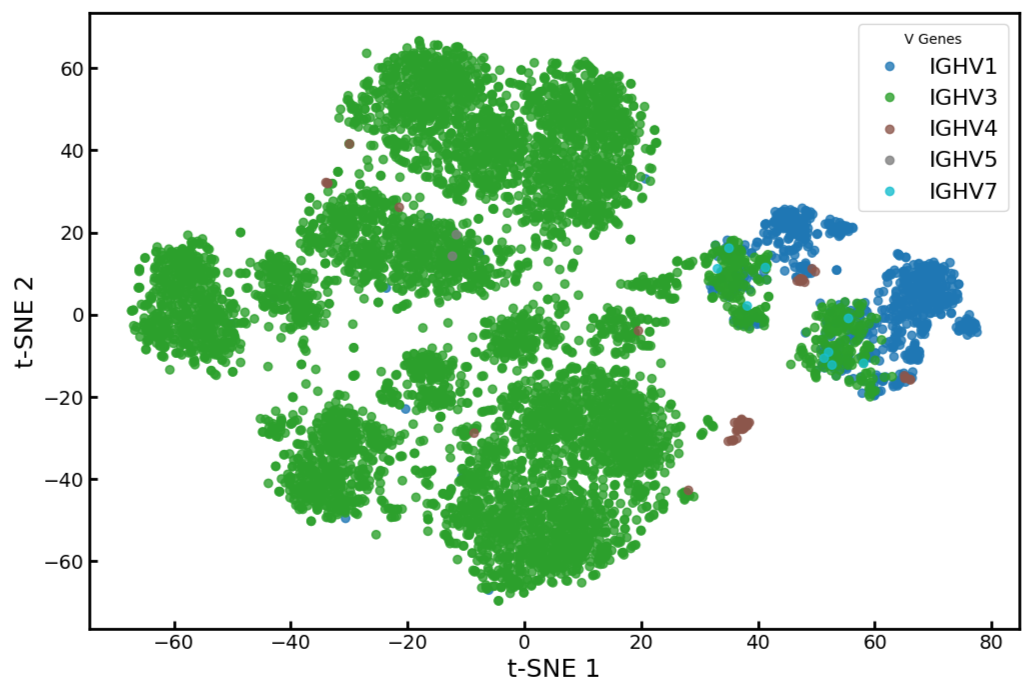}
        \label{fig:tsne v-gene}
    \end{subfigure}
    \hfill
    \begin{subfigure}[t]{0.495\linewidth}
        \caption[]{}
        \includegraphics[width=\linewidth]{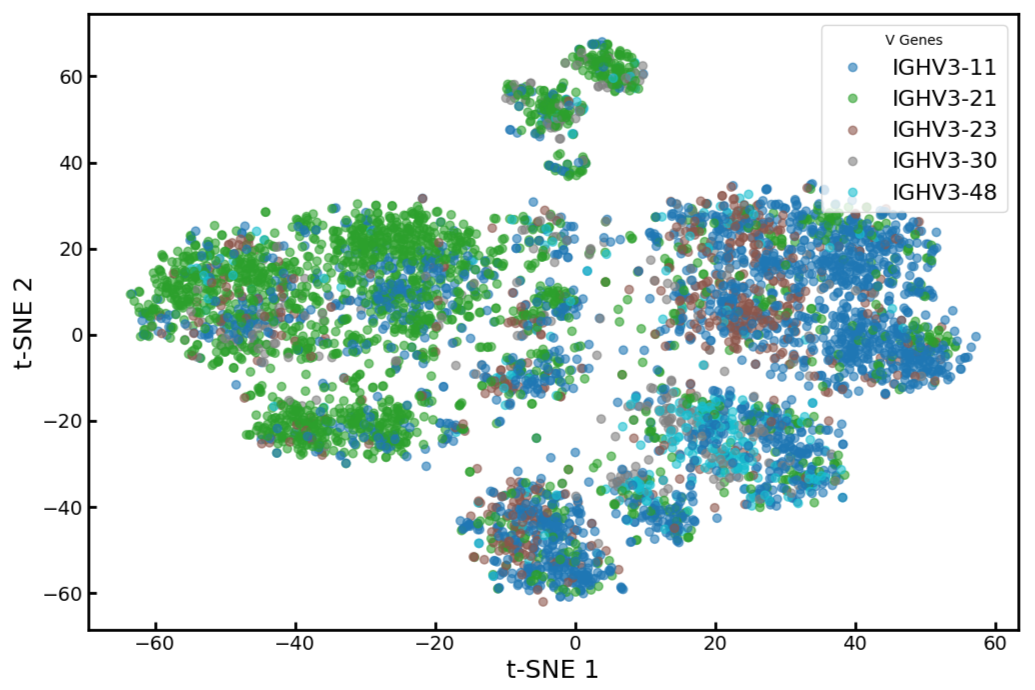}
        \label{fig:tsne v-gene allele}
    \end{subfigure}
    \caption{The diversity in the generated BCR sequences. (a) Length distributions of the generated sequences across different V-J gene families show diversity. (b) t-SNE visualization of generated and training sequences shows distinct differences between clusters, yet not too dissimilar from the natural BCRs. (c) t-SNE visualization of different V-gene families. The population of the IGHV3 gene family is proportionally bigger than the other V gene families. (d) t-SNE visualization of IGHV3 by different gene segments suggests that the model can learn deep representations of the BCR space.}
    \label{fig:diversity}
\end{figure}

A diverse BCR repertoire is crucial for the immune system to recognize and neutralize a wide array of pathogens. To assess AbGPT's capability to generate such diversity, we focused on the CDR-H3 region. The length distributions grouped by various V-J combinations are shown in Figure \ref{fig:length dist}. The variability in CDR-H3 length across different V-J gene families indicates that AbGPT can generate a broad range of sequences, which is critical for enabling diverse antigen recognition. To further evaluate the novelty and biological relevance of the sequences generated by AbGPT, we employed t-SNE visualization to explore the distribution and relationships between the synthetic BCR library and the training data. As shown in Figure \ref{fig:tsne train and gen}, the generated sequences form distinct clusters that are separate from the training data. This suggests that AbGPT is capable of producing novel sequences that do not merely replicate the training data, but maintain essential biological representations. The slight overlap and proximity of these clusters to the training data indicate that AbGPT preserves key biological characteristics that align with natural BCR sequences. Figure \ref{fig:tsne v-gene} presents the t-SNE plot of the top five V-gene family classes observed in the generated BCR library. A predominant representation of IGHV3 is evident, likely influenced by the generation and filtering approaches employed in this study. Specifically, our approach favored sequences with the most frequently observed starting amino acid residues in the training data and filtered out those below a defined perplexity score threshold. This bias may have excluded some viable sequences from other V-gene families. Further analysis of the training data revealed an inherent imbalance in V-gene family classes, which could have contributed to higher perplexity scores for certain classes. Despite this bias, AbGPT demonstrates proficiency in generating and distinguishing sequences across different V-gene families, as evidenced by the formation of distinct clusters in the t-SNE plot. 
To determine whether AbGPT captures deeper biological representations, such as specific gene segments within a V-gene family, we plotted another t-SNE using the V-gene gene segments of the IGHV3 family, the largest V-gene family in our generated BCR library (\ref{fig:tsne v-gene allele}). The clear separation of clusters by gene segments within the IGHV3 family suggests that AbGPT is capable of understanding deep biological hierarchy and designing BCRs of specific characteristics. These findings collectively demonstrate that the model generation is capable of making diverse and novel BCR repertoire, and also contextualizing deep representations that could hugely benefit downstream tasks such as designing antibodies of specific gene classes. However, the observed bias towards certain V-gene families highlights an area for future improvement to create a more balanced representation of the synthetic BCR library.

\subsection{Generated BCRs closely resemble natural antibodies}

\begin{figure}[h] 
    \centering
    \begin{subfigure}[t]{0.49\linewidth}
        \caption[]{}
        \includegraphics[width=\linewidth]{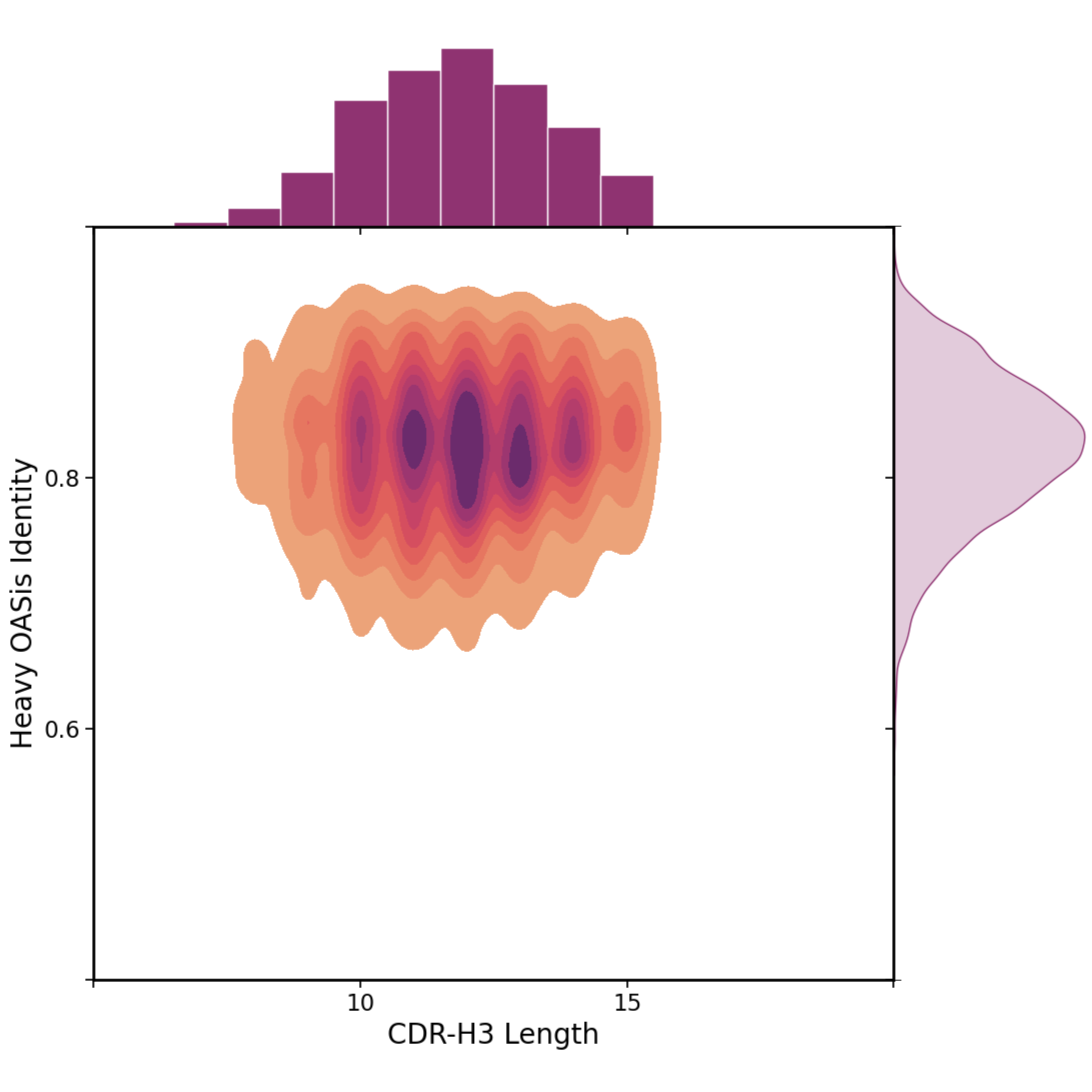}
        \label{fig:human identity}
    \end{subfigure}
    \hfill
    \begin{subfigure}[t]{0.49\linewidth}
        \caption[]{}
        \includegraphics[width=\linewidth]{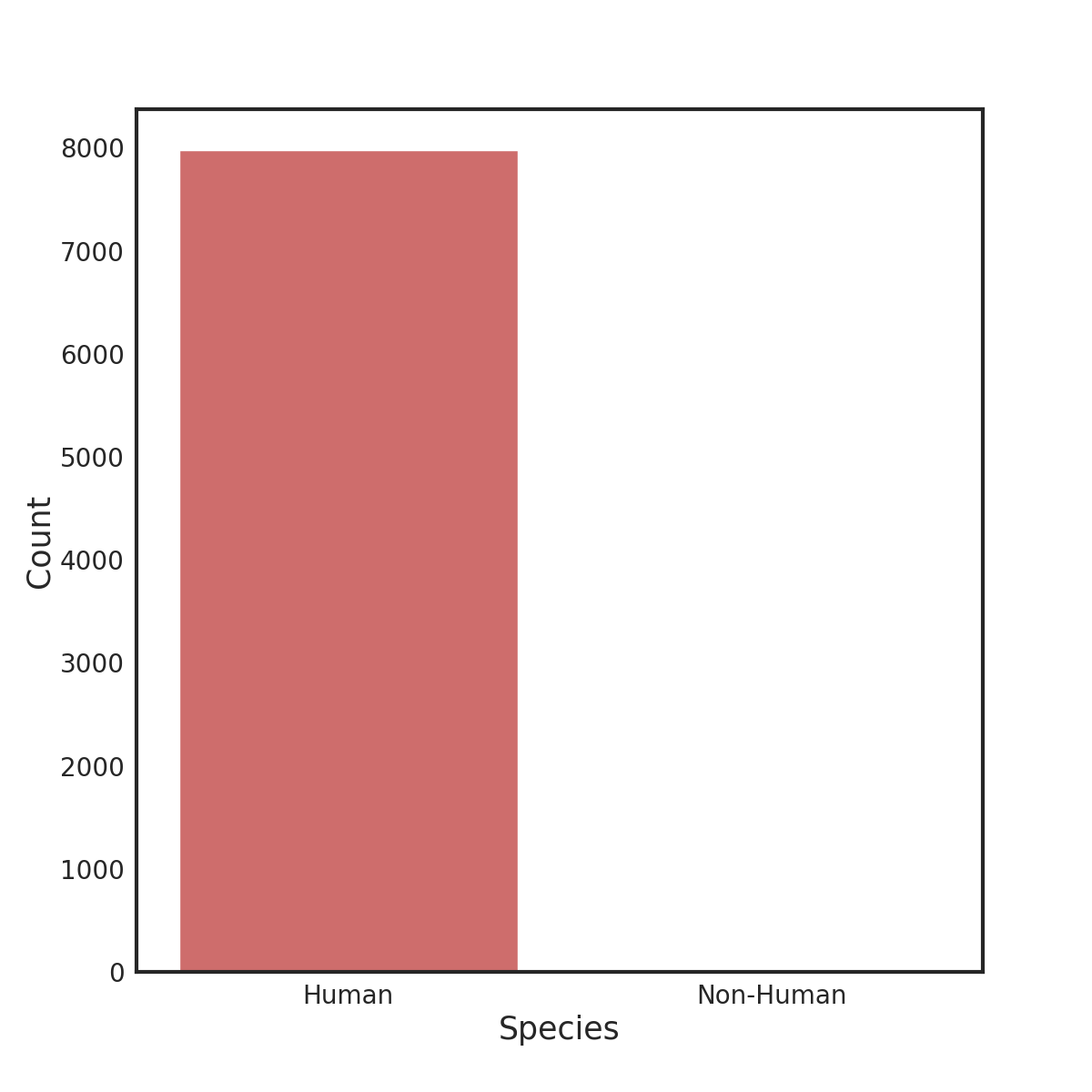}
        \label{fig:human species}
    \end{subfigure}
    \caption{The humanness evaluation of generated BCR sequences. (a) Density plot between Heavy OASis Identity and CDR-H3 sequence length, with prevalence threshold set at relaxed ($\geq$10\% subjects). (b) Species classification histogram validates the sequence generation is 100\% human species.}
    \label{fig:humanness}
\end{figure}

Ensuring that the generated BCR sequences closely resemble natural antibodies is crucial for their potential immunogenicity. Human-like BCR sequence design is essential for proper folding and binding specificity, which ultimately boosts the safety and effectiveness of the generated BCRs. AbGPT excels at producing BCR sequences that exhibit high human likeness, as demonstrated by our analysis in Figure \ref{fig:human identity}. Specifically, our analysis revealed that the generated sequences predominantly display a Heavy OASis Identity metric of approximately 0.8, demonstrating a high degree of resemblance to natural human antibodies at a prevalence threshold of $\geq$ 10 \% of human subjects. This metric measures the fraction of heavy chain 9-mer peptides classified as human, with a value of 1.0 indicating the highest degree of human likeness. The prevalence threshold determines the percentage of human subjects in the Observed Antibody Space database that must contain a given peptide for it to be classified as human. 
The analysis demonstrates no direct correlation between the length of the CDR-H3 sequence and the OASis human-likeness evaluation, suggesting that the quality of BCR sequence design by the model does not depend on the sequence length. Additionally, species classification using ANARCI further validated the species origin of our generated sequences, as shown in Figure \ref{fig:human species}. Together, these results confirm that the model produces BCR sequences that preserve the properties of natural human immunoglobulins. By closely aligning with human-like characteristics, this BCR library can facilitate the development of safe and effective therapeutic antibodies.

\section{Conclusion}
In this study, we introduced AbGPT, a generative model fine-tuned on 57 million natural human BCR sequences using ProtGPT2, a model trained across the entire protein space. Our approach utilized context prompting with four amino acids to differentiate between heavy and light chain sequences and employed a robust filtering process to create a synthetic BCR library. Through this methodology, we successfully generated over 15,000 BCR sequences exhibiting high developability profiles, including properties such as solubility, aggregation propensity, structural stability, and humanness. Our results demonstrate that AbGPT is capable of advancing the design of full-length BCR sequences with a diverse and human-like repertoire, essential for therapeutic applications. Despite these successes, our approach is not without limitations. The diversity of the synthetic BCR library is currently constrained by our generation and filtering framework. This potentially limits the exploration of the full antibody space. Furthermore, the observed lower structure prediction scores in certain CDR-H3 sequences suggest that further investigation is needed to identify motifs and hypervariable regions that could guide targeted modifications for optimization. Future work should focus on addressing these limitations by adopting multimodal approaches that integrate antibody-antigen interaction data, which could improve the accuracy of binding affinity predictions and broaden the design space. Exploring weighted attention mechanisms on different segments of a BCR sequence within this framework may also help in optimizing specific regions, particularly those that contribute to structural stability and functional efficacy. Another promising direction is the development of a multi-LLM agent framework, where specialized language model agents, each fine-tuned for specific domain knowledge such as sequence generation, structural prediction, and interaction modeling, can collaborate to create a more holistic design process. To facilitate further advancements in the field, we have made our model and associated scripts publicly available. We believe that our contributions will serve as a valuable resource for the community and provide a foundation for future developments in antibody design.

\section{Data and software availability}

The necessary code used for model training and sequence generation in this study can be accessed here:
\href{https://github.com/deskk/AbGPT}{https://github.com/deskk/AbGPT}

\begin{acknowledgement}
We express our gratitude to the authors of ProtGPT2 for their model, fine-tuning, and generation pipeline. We also thank the authors of AntiBERTa's authors for their contributions to the field and the fine-tuning dataset. Additionally, we acknowledge Brian Loyal for his efforts in processing the fine-tuning dataset. We are grateful of the insightful discussions from our colleagues who greatly contributed to this work.

\end{acknowledgement}

\newpage
\bibliography{reference}

\newpage
\section{Appendix}
\subsection{Model Implementation and Optimal Hyperparameters}

AbGPT was fine-tuned using optimal hyperparameters to maximize generative performance. The training was conducted over 5 epochs with a learning rate of $1 \times 10^{-6}$.  The Adam optimizer was employed with parameters $\beta_1 = 0.9$, $\beta_2 = 0.999$, and $\epsilon = 1 \times 10^{-8}$. A linear learning rate scheduler was used throughout the training process. For tokenization, the Byte-Pair Encoding (BPE) tokenizer was utilized. We used Nvidia Lambda RTX A6000 (48GB GDDR6 memory) for this work. 

\begin{table}[h]
\centering
\begin{tabular}{p{6cm} p{4cm}} 
\toprule
\textbf{Number of layers} & 36 \\
\textbf{Embedding dimension} & 1280 \\
\textbf{Hidden dimension} & 1280 \\
\textbf{Feed-forward dimension} & 5120 \\
\textbf{Attention heads} & 20 \\
\textbf{Total parameters} & 734M \\
\bottomrule
\end{tabular}
\caption{Model Hyperparameters}
\label{model implementation}
\end{table}

\subsection{Sequence Generation Details}
\begin{figure}[t!]
     \centering
     \includegraphics[width=0.8\linewidth]{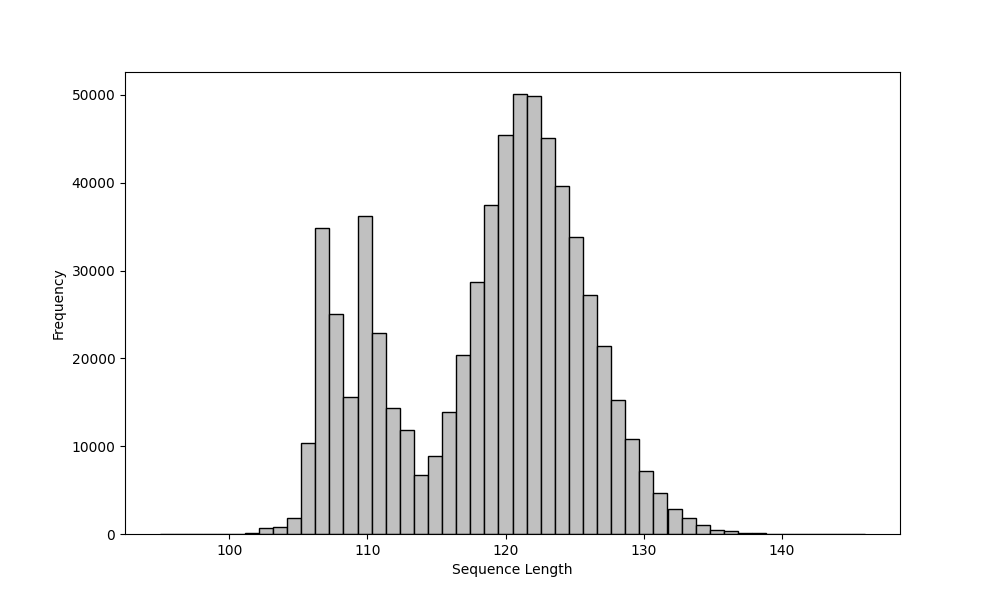}
     \caption{Sequence length distribution of the training dataset.} 
     \label{fig:histogram}
\end{figure}

We opted for generation of BCR heavy and light chains without utilizing any conditional tags because BCR heavy and light chains both have unique starting amino acids. Additionally, analyzing the length distribution of our training dataset revealed a distinct bimodal distribution, which we hypothesize reflects the distribution of heavy and light chains in the BCR repertoire, as shown in Figure \ref{fig:histogram}. The pipeline is designed to generate sequences for both light and heavy chains of BCRs. We opted for 4 starting amino acids as preliminary experimentation in Table \ref{prompt experiment} shows lower perplexity score in sequence generation. We utilize most frequent starting residues for each chain type, as shown in Table \ref{prompt table}. For each chain type, the pipeline generates sequences in batches of 100. This batch generation process is repeated 1000 times, resulting in a total of 100,000 sequences per starting residue. The minimum sequence length is set to 20 tokens for light chains and 28 tokens for heavy chains, with each token approximately representing 4 amino acids. The sequences generated in each batch are cleaned and filtered to meet length constraints specific to the chain type: light chains are filtered to be between 100 and 120 amino acids, and heavy chains between 110 and 140 amino acids. After filtering, a sample of 10,000 sequences is randomly selected from the filtered list. Additionally, sequences containing 'X' or 'B' were removed. The second step involves further filtering based on sequence quality. Only sequences with a perplexity score below a threshold of 13.0 are retained.

\begin{table}[h]
\centering
\begin{tabular}{p{7cm} p{5cm}} 
\toprule
\textbf{Number of Starting Amino Acids} & \textbf{Perplexity Score Range} \\
\midrule
$<|endoftext|>$ & 20 - 351 \\
$<|endoftext|>$ Q & 17 - 147 \\
$<|endoftext|>$ QVQ & 14 - 268 \\
$<|endoftext|>$ QVQL & 5 - 36 \\
\bottomrule
\end{tabular}
\caption{Experimentation of prompting strategy indicates low perplexity score when 4 amino acids are used}
\label{prompt experiment}
\end{table}

\begin{table}[h]
\centering
\begin{tabular}{p{5cm} p{8cm}} 
\toprule
\textbf{Chain Type} & \textbf{Starting Residues} \\
\midrule
Light Chain & DIQM, EIVL, QSAL, QSVL, EIVM \\
Heavy Chain & QVQL, EVQL, VQLV, QLQL, QSGA\\
\bottomrule
\end{tabular}
\caption{Starting residue amino acids for light and heavy chains.}
\label{prompt table}
\end{table}

\subsection{Imbalance V-gene Population of Training Data}
Here we include a 2-dimensional t-SNE plot of the training data clustered by V-gene identity. This visualization highlights the insufficiency of certain V-gene families within the dataset. Our findings support the hypothesis that V-gene families with higher perplexity scores are underrepresented in the generated sequences. Consequently, this results in a smaller number of sequences for these specific V-gene families.

\begin{figure*}[htp] 
    \centering
        \includegraphics[width=\linewidth]{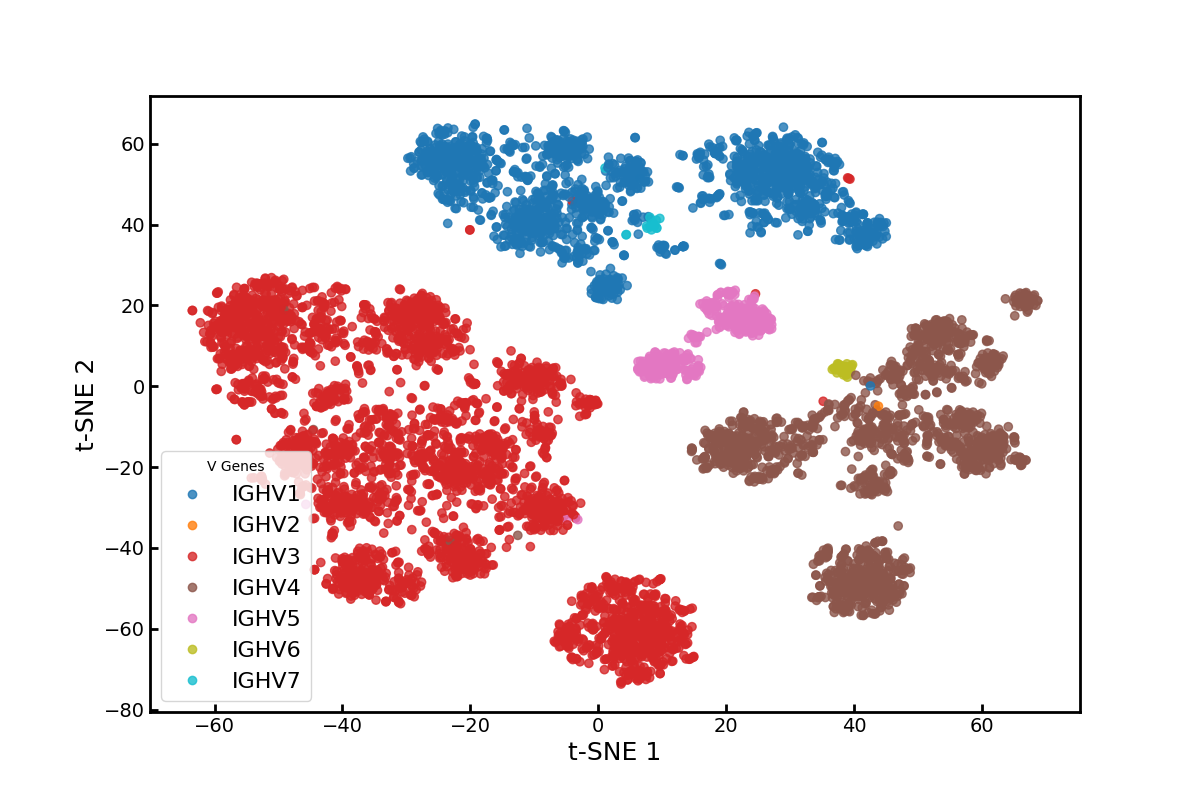}
        \label{fig:tsne}
    \caption{The t-SNE plot of V-gene families sampled from natural BCR repertoire training data.}
\end{figure*}

\end{document}